\documentclass[onecolumn,aps,prc,showpacs,superscriptaddress]{revtex4}
\usepackage{epsfig}

\newenvironment{color}[3]
{
}{
}

\newcommand {\mean}[1]	{\langle#1\rangle}
\newcommand {\cabp}	{\cos(\alpha+\beta-2\psi)}
\newcommand {\cabc}	{\cos(\alpha+\beta-2c)}
\newcommand {\cab}	{\cos(\alpha-\beta)}
\newcommand {\pT}	{p_{T}}
\newcommand {\vc}	{v_{2,c}}
\newcommand {\vcl}	{v_{2,{\rm clust}}}
\newcommand {\vbg}	{v_{2,{\rm bkgd}}}
\newcommand {\dphi}	{\Delta\phi}
\newcommand {\phid}	{\phi_{\Delta}}
\newcommand {\etad}	{\eta_{\Delta}}
\newcommand {\xUS}	{x_{_{US}}}
\newcommand {\xLS}	{x_{_{LS}}}
\newcommand {\wSA}	{w_{_{SA}}}
\newcommand {\wBB}	{w_{_{BB}}}
\newcommand {\wwSA}	{w_{_{2,SA}}}
\newcommand {\vSA}	{v_{_{2,SA}}}
\newcommand {\vBB}	{v_{_{2,BB}}}
\newcommand {\Ncl}	{N_{\rm clust}}
\newcommand {\Mcl}	{M_{\rm clust}}
\newcommand {\phicl}	{\phi_{\rm clust}}
\newcommand {\acc}	{\mathcal{A}(\etad)}
\newcommand {\Npart}	{N_{\rm part}}
\newcommand {\meas}	{{\rm meas}}

\newcommand {\be}	{\begin{equation}}
\newcommand {\ee}	{\end{equation}}
\newcommand {\bea}	{\begin{eqnarray}}
\newcommand {\eea}	{\end{eqnarray}}

\begin{document}

\title{Effects of Cluster Particle Correlations on Local Parity Violation Observables}
\author{Fuqiang Wang}
\affiliation{Department of Physics, Purdue University, 525 Northwestern Ave., West Lafayette, IN 47907}

\begin{abstract}
We investigate effects of cluster particle correlations on two- and three-particle azimuth correlator observables sensitive to local strong parity violation. We use two-particle angular correlation measurements as input and estimate the magnitudes of the effects with straightforward assumptions. We found that the measurements of the azimuth correlator observables by the STAR experiment can be entirely accounted for by cluster particle correlations together with a reasonable range of cluster anisotropy in non-peripheral collisions. Our result suggests that new physics, such as local strong parity violation, may not be required to explain the correlator data.
\end{abstract}


\maketitle

\section{Introduction}

Relativistic heavy-ion collisions at RHIC have created a hot and dense medium that exhibits properties of a strongly coupled Quark-Gluon Plasma (sQGP)~\cite{wps}. Approximate chiral symmetry may be restored in the sQGP. It was recently suggested that metastable domains may form in such sQGP state where the parity and time-reversal symmetries are locally violated~\cite{PV,PVquench}. Such violation would lead to separation of positive and negative particles due to the chiral magnetic effect along the system's orbital angular momentum into the two hemispheres separated by the reaction plane~\cite{PV,PVquench}. 

The most direct consequence of this charge separation is a negative correlation of multiplicity asymmetry of positive particles between the hemispheres separated by the reaction plane and that of negative particles~\cite{PV}. Such a negative correlation is in addition to any (background) correlations that may exist due to other dynamics of the collision, the magnitude of which may be assessed by correlation of multiplicity asymmetries between the hemispheres separated by the plane normal to the reaction plane.

Another consequence of the charge separation is a positive correlator $\mean{\cabp}$ of unlike-sign (US) particle pairs and a negative $\mean{\cabp}$ of like-sign (LS) particle pairs, where $\alpha$ and $\beta$ are the azimuthal angles of the two particles and $\psi$ is the reaction plane azimuth~\cite{Voloshin}. The reaction plane azimuthal angle is, however, not fixed but random in heavy-ion collisions. In order to estimate the reaction plane angle, a third particle, $c$, may be used to correlate with $\alpha$ and $\beta$, correcting for the resolution effect ($\vc$, the elliptic flow of particle $c$)~\cite{Voloshin}. Namely
\be
\mean{\cabp}\approx\mean{\cabc}/\vc.
\label{eq1}
\ee
This assumes three-particle correlation is negligible. 

The three-particle azimuthal correlator $\mean{\cabc}$ has been measured and is used to  deduce the two-particle azimuthal correlator $\mean{\cabp}$ by the STAR experiment~\cite{STAR}. The measurements show a negative correlator $\mean{\cabp}$ for LS pairs and a small, close to zero, correlator for US pairs~\cite{STAR}. The LS pair result is qualitatively consistent with the expectation from local strong parity violation~\cite{PV,PVquench}. The US pair result, however, is inconsistent with the initial expectation where the US and LS pair correlations should be equal in magnitude and opposite in sign~\cite{PV}. To explain the preliminary version of the STAR data~\cite{Selyuzhenkov}, it was suggested that at least one of the particles from a back-to-back US pair from local parity violation would have to traverse and interact with the medium and its angular correlation with the other particle of the pair would be significantly reduced~\cite{PVquench}. In fact, in this medium interaction scenario, correlations between particle pairs from local parity violation domains formed in the interior of the collision medium would be lost, and only those from pairs emitted from the surface could survive. In other words, the LS pair correlation is due to those pairs from local parity violation domains on the surface, and the back-to-back US pair correlation is lost~\cite{PVquench}.

The three-particle correlator observable is parity-even and is subject to background correlations that are reaction-plane-dependent, some of which are discussed in detail in Ref.~\cite{STAR}. This is easy to see in the following extreme: a small opening angle pair perpendicular to the reaction plane is indistinguishable from a back-to-back pair parallel to the reaction plane in the correlator variable $\cabp$, and vice versa. More modestly, particle correlations from clusters which themselves possess anisotropy can give rise to observable signals in $\mean{\cabp}$. Cluster correlations can have different effects on LS and US pairs, because LS and US contributions from clusters are likely different, such as from jet-correlations~\cite{Back2back}.

In this paper, we investigate quantitatively effects of cluster particle correlations on the azimuth correlator observables $\mean{\cabc}$ and $\mean{\cabp}$. We first present analytical results. We then use experimental measurements of two-particle angular correlations~\cite{Daugherity,STAR} as input to estimate effects of cluster particle correlations on the correlator observables. Since we do not have experimental information on cluster anisotropy, we calculate how much cluster anisotropies are needed in order to fully account for the correlator measurements~\cite{STAR}. We then judge the plausibility of the needed cluster anisotropies to either confirm or disprove cluster correlations as a possible explanation for the correlator measurements.

\section{The Cluster Model and Results}

Assume events are composed of hydrodynamic-like particles plus small-angle (SA, $|\dphi|<\pi/2$) clusters and back-to-back (BB, $|\dphi|>\pi/2$) clusters. We will use operational definitions of SA-cluster to be composed of one or more small-angle particle pairs, and BB-cluster to be composed of one or more back-to-back particle pairs. Note, with these operational definitions, that a conventional back-to-back cluster of $a$ particles on one side and $b$ particles on the other side is made of two SA-clusters (with numbers of particle pairs of $a^2$ and $b^2$) and one BB-cluster (with number of particles $2ab$). Here we have taken $a$ and $b$ to be large for simplicity, and we will assume Poisson statistics for particle multiplicity in clusters. Also note that not all SA-clusters have a back-side partner. The anisotropies of SA- and BB-clusters can therefore be different. 

Now consider US and LS particle pairs from clusters. They can come from either SA-clusters (i.e. SA particle pairs) or BB-clusters (i.e. BB particle pairs). The relative fractions of particle pairs from SA-clusters and BB-clusters can be different for US-pairs and LS-pairs. Let $\xUS$ be the fraction of US-pairs from SA-clusters (and $1-\xUS$ the fraction from BB-clusters), and $\xLS$ be the fraction of US-pairs from SA-clusters (and $1-\xLS$ the fraction from BB-clusters). We shall first obtain $\xUS$ and $\xLS$ from two-particle angular correlation measurements.

The following two-particle correlators are measured for US and LS pairs~\cite{STAR}:
\bea
\mean{\cab}_{_{US}}&=&\xUS\mean{\cab}_{_{SA}}+(1-\xUS)\mean{\cab}_{_{BB}}=\xUS\wSA-(1-\xUS)\wBB,\label{eq2a}\\
\mean{\cab}_{_{LS}}&=&\xLS\mean{\cab}_{_{SA}}+(1-\xLS)\mean{\cab}_{_{BB}}=\xLS\wSA-(1-\xLS)\wBB,\label{eq2b}
\eea
where 
\bea
\wSA&\equiv&\mean{\cab}_{_{SA}},\label{eq3a}\\
\wBB&\equiv&-\mean{\cab}_{_{BB}}\label{eq3b}
\eea
are the average angular spread of particle pairs from SA- and BB-clusters, respectively. In Eqs.~(\ref{eq2a}) and (\ref{eq2b}) we have taken the two-particle back-to-back correlations to be the same between US and LS pairs. We have assumed in Eqs.~(\ref{eq2a}) and (\ref{eq2b}) that the SA two-particle azimuthal correlations to be of the same shape for US and LS pairs. This is a reasonable assumption because same-side correlations of US and LS pairs have similar shapes although their magnitudes are different, e.g. in jet-like correlations~\cite{Back2back}.

Information about cluster particle pairs can be obtained from two-particle azimuthal correlations. STAR has measured two-particle correlations integrated over transverse momentum ($\pT$), in $(\etad,\phid)$, the two-particle pseudorapidity and azimuth differences~\cite{Daugherity}. 
The correlation functions are parameterized by the sum of a near-side Gaussian, a negative dipole, and a quadrupole corresponding to elliptic flow~\cite{Daugherity}. The sum of the two former terms is considered to be correlations due to clusters. It is given by~\cite{Daugherity}: 
\be
\frac{d^2N}{d\phid d\etad}=\frac{V_0}{\sqrt{2\pi}\sigma}\exp\left(-\frac{\phid^2}{2\sigma^2}\right)G(\etad)-A_{\phid}\cos\phid.
\label{eq4}
\ee
Here $G(\etad)$ is a Gaussian in $\etad$ normalized to unity that is of no interest in our study. The first term in Eq.~(\ref{eq4}) r.h.s.~is the near-side Gaussian and the second term is the negative dipole. We can obtain the SA pair azimuthal spread as
\be
\wSA\equiv\mean{\cos\phid}_{_{SA}}
=\frac{\int_{-\pi/2}^{\pi/2}\left[\frac{V_0}{\sqrt{2\pi}\sigma}\exp\left(-\frac{\phid^2}{2\sigma^2}\right)G(\etad)-A_{\phid}\cos\phid\right]\cos\phid d\phid\acc d\etad}{\int_{-\pi/2}^{\pi/2}\left[\frac{V_0}{\sqrt{2\pi}\sigma}\exp\left(-\frac{\phid^2}{2\sigma^2}\right)G(\etad)-A_{\phid}\cos\phid\right]d\phid\acc d\etad}
=\frac{Ve^{-\sigma^2/2}-\pi A_{\phid}}{V-4A_{\phid}},
\label{eq5}
\ee
where $\acc$ is the two-particle $\etad$ acceptance of the STAR detector, and $V$ is the integrated volume within the acceptance (which is not equal to $V_0$)~\cite{Daugherity,Wang}. The extracted $\wSA$ ranges from 0.58 in peripheral Au+Au collisions to 0.85 in central collisions. For $\wBB$ we take the width of the away-side dipole: 
\be
\wBB\equiv -\mean{\cos\phid}_{_{BB}}=\frac{\int_{\pi/2}^{3\pi/2}\left(A_{\phid}\cos\phid\right)\cos\phid d\phid}{\int_{\pi/2}^{3\pi/2}A_{\phid}\cos\phid d\phid}=\frac{\pi}{4}.
\label{eq6}
\ee
The $\wBB$ is a fixed value because the away-side correlation shape can be satisfactorily described by the same functional form of a negative dipole.

It is worthwhile to note that the cluster shape quantities are extracted from the measured angular correlations, thus they are immune to the underlying physics mechanisms generating those correlations. Besides cluster correlations, a negative dipole can be also generated, for example, by the statistical global momentum conservation. It was estimated, however, that the global momentum conservation effect is significantly smaller than the measured dipole strength. On the other hand, in order to estimate the cluster size that will be needed for our study below, some production mechanisms for the clusters have to be assumed. We will discuss those assumptions later.


The pair quantities in Eqs.~(\ref{eq2a}) and (\ref{eq2b}) measured in experiment are diluted by hydro-like particles (those pair quantities are all zero for hydro-like particle pairs as well as cross-pairs of hydro-like and cluster particles):
\bea
\mean{\cab}_{_{US}}^{\meas}&=&f_{_{2,US}}\mean{\cab}_{_{US}},\label{eq7a}\\
\mean{\cab}_{_{LS}}^{\meas}&=&f_{_{2,LS}}\mean{\cab}_{_{LS}}.\label{eq7b}
\eea
The dilution factors are 
\bea
f_{_{2,US}}&=&N_{_{US}}/(N^2/2),\label{eq8a}\\
f_{_{2,LS}}&=&N_{_{LS}}/(N^2/2),\label{eq8b}
\eea
for US and LS pairs (numbers of pairs $N_{_{US}}$ and $N_{_{LS}}$), respectively, where we have assumed the total numbers of US and LS pairs are equal in the event ($N$ is total particle multiplicity). The total number of cluster particle pairs is
\be
N_{_{US}}+N_{_{LS}}=\Ncl\Mcl^2,\label{eq9}
\ee
where $\Ncl$ is the number of clusters and $\Mcl$ is the particle multiplicity per cluster (cluster size). The numbers of US and LS pairs from {\em clusters} are not necessarily equal. Since there is no charge-sign difference in the back-to-back particle pair correlations:
\be
(1-\xUS)N_{_{US}}=(1-\xLS)N_{_{LS}},\label{eq10}
\ee
we obtain the dilution factors:
\bea
f_{_{2,US}}&=&\frac{1-\xLS}{1-(\xUS+\xLS)/2}f_2,\label{eq11a}\\
f_{_{2,LS}}&=&\frac{1-\xUS}{1-(\xUS+\xLS)/2}f_2,\label{eq11b}
\eea
with
\be
f_2=\Ncl\Mcl^2/N^2\label{eq11c}.
\ee
Note $\Ncl\Mcl^2/N$ --the number of correlated pairs per charged particle-- is a measured quantity in two-particle correlation function; it is obtained by integrating the correlation function over the measured acceptance (which is the $V$ in Eq.~(\ref{eq4}), $V=\Ncl\Mcl^2/N$)~\cite{Daugherity,Wang}. The measured $V$ ranges from 0.3 in peripheral Au+Au collisions to 1.9 in central collisions. The measured $\Ncl\Mcl^2$ is not affected by the individual quantities of the cluster size $\Mcl$ or the number of clusters $\Ncl$ which are not experimentally measured.

We can now obtain $\xUS$ and $\xLS$ from the measured $\mean{\cab}_{_{US}}^{\meas}$ and $\mean{\cab}_{_{LS}}^{\meas}$. With short notations:
\bea
t_{_{US}}&=&\mean{\cab}_{_{US}}^{\meas}/f_2,\label{eq12a}\\
t_{_{LS}}&=&\mean{\cab}_{_{LS}}^{\meas}/f_2,\label{eq12b}
\eea
Eqs.~(\ref{eq2a}) and (\ref{eq2b}) become
\bea
t_{_{US}}[1-(\xUS+\xLS)/2]&=&\xUS(1-\xLS)\wSA-(1-\xUS)(1-\xLS)\wBB,\label{eq13a}\\
t_{_{LS}}[1-(\xUS+\xLS)/2]&=&\xLS(1-\xUS)\wSA-(1-\xUS)(1-\xLS)\wBB.\label{eq13b}
\eea
By simple algebra, we have
\be
\xUS^2-\frac{\wSA^2+\wSA(3t_{_{US}}-t_{_{LS}})/2+2\wSA\wBB+\wBB(t_{_{US}}-t_{_{LS}})}{(\wSA+\wBB)(\wSA+(t_{_{US}}-t_{_{LS}})/2)}\xUS+\frac{\wSA\wBB+\wSA t_{_{US}}+\wBB(t_{_{US}}-t_{_{LS}})/2}{(\wSA+\wBB)(\wSA+(t_{_{US}}-t_{_{LS}})/2)}=0,\label{eq14a}
\ee
\be
\xLS^2-\frac{\wSA^2+\wSA(3t_{_{LS}}-t_{_{US}})/2+2\wSA\wBB+\wBB(t_{_{LS}}-t_{_{US}})}{(\wSA+\wBB)(\wSA+(t_{_{LS}}-t_{_{US}})/2)}\xLS+\frac{\wSA\wBB+\wSA t_{_{LS}}+\wBB(t_{_{LS}}-t_{_{US}})/2}{(\wSA+\wBB)(\wSA+(t_{_{LS}}-t_{_{US}})/2)}=0.\label{eq14b}
\ee
From Eqs.~(\ref{eq14a}) and (\ref{eq14b}) we can solve for $\xUS$ and $\xLS$.

Table~\ref{tab} shows the obtained fractions $\xUS$ and $\xLS$ given cluster particle correlation inputs from two-particle angular correlations~\cite{Daugherity} and two-particle correlator quantities $\mean{\cab}_{_{US}}^{\meas}$ and $\mean{\cab}_{_{LS}}^{\meas}$ measured by STAR~\cite{STAR}. 

\subsection{Effect of three-particle correlation from clusters\label{sec2A}}

STAR has measured three-particle correlators for US and LS pairs, $\alpha$ and $\beta$, with a third particle $c$ regardless of its charge sign. It is assumed that three-particle correlation is negligible, so particle $c$ can be used as a single-particle estimator of the reaction plane to obtain the two-particle correlators from the three-particle correlator measurements by Eq.~(\ref{eq1}). One supporting evidence for the assumption comes from the consistent results of $\mean{\cabp}\approx\mean{\cabc}/\vc$ using particle $c$ from the main Time Projection Chamber (TPC) or the forward TPC's while the particle pairs (US and LS) $\alpha$ and $\beta$ come from the main TPC~\cite{STAR}. However, it is possible that the probability for a triplet to be correlated may drop with the pseudorapidity gap between the particle $c$ and the other two particles in the main TPC, in a similar way to the $\vc$ dependence on pseudorapidity~\cite{PHOBOS}.

If large clusters exist, as suggested by low-$\pT$ two-particle angular correlation measurements~\cite{Daugherity}, then finite three-particle correlation should exist. We shall estimate the effect of three-particle correlation using two-particle correlations~\cite{Daugherity}. Consider particle triplets from the same cluster, where SA and BB still stand for a pair of $\alpha$ and $\beta$, and the third particle $c$ can be on either side.
\bea
\mean{\cabc}_{_{SA}}&=&\mean{\cos(\dphi_\alpha+\dphi_\beta+2\phicl-2c)}_{_{SA}}=\mean{\cos(\dphi_\alpha+\dphi_\beta+2\dphi_c)}_{_{SA}},\label{eq15a}\\
\mean{\cabc}_{_{BB}}&=&\mean{\cos(\dphi_\alpha+\dphi_\beta+2\phicl-2c)}_{_{BB}}=\mean{\cos(\dphi_\alpha+\dphi_\beta+2\dphi_c)}_{_{BB}}.\label{eq15b}
\eea
Here $\dphi=\phi-\phicl$ is the particle azimuth relative to the cluster axis, $\phicl$. In Eq.~(\ref{eq15a}), which side the particle $c$ is does not really matter (because the angle is $2c$). In Eq.~(\ref{eq15b}), the particle $c$ is on the same side of either $\alpha$ or $\beta$, and thereby either $\dphi_\alpha$ or $\dphi_\beta$ is larger than $\pi/2$. Assuming that emission of daughter particles in clusters is independent of each other, we can obtain
\bea
\mean{\cabc}_{_{SA}}&\approx&\mean{\cos\dphi_\alpha}_{_{SA}}^2\mean{\cos2\dphi_c}_{_{SA}}\approx\mean{\cos(\dphi_\alpha-\dphi_\beta)}_{_{SA}}\mean{\cos2\dphi_c}_{_{SA}}=\wSA\wwSA,\label{eq15c}\\
\mean{\cabc}_{_{BB}}&\approx&\mean{\cos\dphi_\alpha}_{_{BB}}^2\mean{\cos2\dphi_c}_{_{SA}}\approx\mean{\cos(\dphi_\alpha-\dphi_\beta)}_{_{BB}}\mean{\cos2\dphi_c}_{_{SA}}=-\wBB\wwSA.\label{eq15d}
\eea
Here $\dphi_\alpha-\dphi_\beta=\phid$ is the azimuth difference used in two-particle correlation measurement, and $\wwSA=\mean{\cos2\dphi}_{_{SA}}$ is the average azimuthal spread of SA-clusters:
\bea
\wwSA&\equiv&\mean{\cos2\dphi}_{_{SA}}\approx\mean{\cos2\phid}^{1/2}\nonumber\\
&=&\left(\frac{\int_{-\pi/2}^{\pi/2}\left[\frac{V_0}{\sqrt{2\pi}\sigma}\exp\left(-\frac{\phid^2}{2\sigma^2}\right)G(\etad)-A_{\phid}\cos\phid\right]\cos2\phid d\phid\acc d\etad}{\int_{-\pi/2}^{\pi/2}\left[\frac{V_0}{\sqrt{2\pi}\sigma}\exp\left(-\frac{\phid^2}{2\sigma^2}\right)G(\etad)-A_{\phid}\cos\phid\right]d\phid\acc d\etad}\right)^{1/2}
=\left(\frac{Ve^{-2\sigma^2}-4A_{\phid}/3}{V-4A_{\phid}}\right)^{1/2}.
\label{eq16}
\eea
Note in Eq.~(\ref{eq15d}), it is the azimuthal spread of SA (not BB) clusters as well because particle $c$ is always on the same side of either particle $\alpha$ or $\beta$.

	We can estimate effects of three-particle correlations in US and LS pairs of particles $\alpha$ and $\beta$ by
\bea
\mean{\cabc}_{_{US}}&=&\xUS\mean{\cabc}_{_{SA}}+(1-\xUS)\mean{\cabc}_{_{BB}}\approx\xUS\wSA\wwSA-(1-\xUS)\wBB\wwSA,\label{eq17a}\\
\mean{\cabc}_{_{LS}}&=&\xLS\mean{\cabc}_{_{SA}}+(1-\xLS)\mean{\cabc}_{_{BB}}\approx\xLS\wSA\wwSA-(1-\xLS)\wBB\wwSA.\label{eq17b}
\eea
Comparing Eqs.~(\ref{eq17a}) and (\ref{eq17b}) to Eqs.~(\ref{eq2a}) and (\ref{eq2b}), we see that 
\bea
\mean{\cabc}_{_{US}}&=&\mean{\cab}_{_{US}}\cdot\wwSA,\label{eq18a}\\
\mean{\cabc}_{_{LS}}&=&\mean{\cab}_{_{LS}}\cdot\wwSA.\label{eq18b}
\eea
The only assumption in arriving at Eqs.~(\ref{eq18a}) and (\ref{eq18b}) is that {\em particle emission azimuths within a cluster are independent of each other}. This is a reasonable assumption when the clusters consist of a relatively large number of particles. Under this assumption, three-particle correlation is completely determined by two-particle correlation. 

These three-particle correlation effects are diluted by hydro-like particles,
\bea
\mean{\cabc}_{_{US}}^{\meas}&=&f_{3,US}\mean{\cabc}_{_{US}},\label{eq19a}\\
\mean{\cabc}_{_{LS}}^{\meas}&=&f_{3,LS}\mean{\cabc}_{_{LS}},\label{eq19b}
\eea
 by a factor of 
\bea
f_{3,US}&=&\frac{1-\xLS}{1-(\xUS+\xLS)/2}f_3,\label{eq20a}\\
f_{3,LS}&=&\frac{1-\xUS}{1-(\xUS+\xLS)/2}f_3,\label{eq20b}
\eea
with
\be
f_3=\Ncl\Mcl^3/N^3\label{eq20c}.
\ee
Again, the measured three-particle correlation is determined by the measured two-particle correlation:
\bea
\mean{\cabc}_{_{US}}^{\meas}&=&\mean{\cab}_{_{US}}^{\meas}\cdot\wwSA\cdot\Mcl/N,\label{eq21a}\\
\mean{\cabc}_{_{LS}}^{\meas}&=&\mean{\cab}_{_{LS}}^{\meas}\cdot\wwSA\cdot\Mcl/N.\label{eq21b}
\eea
The only inputs for this determination are the near-side angular spread $\wwSA$ by Eq.~(\ref{eq16}) which is well measured and the cluster size $\Mcl$ which can be estimated. We estimate $\Mcl$ from the measured $\Ncl\Mcl^2$ assuming {\em binary scaling for the number of clusters, $\Ncl$}.

\begin{figure}
\begin{center}
\includegraphics[width=0.45\textwidth]{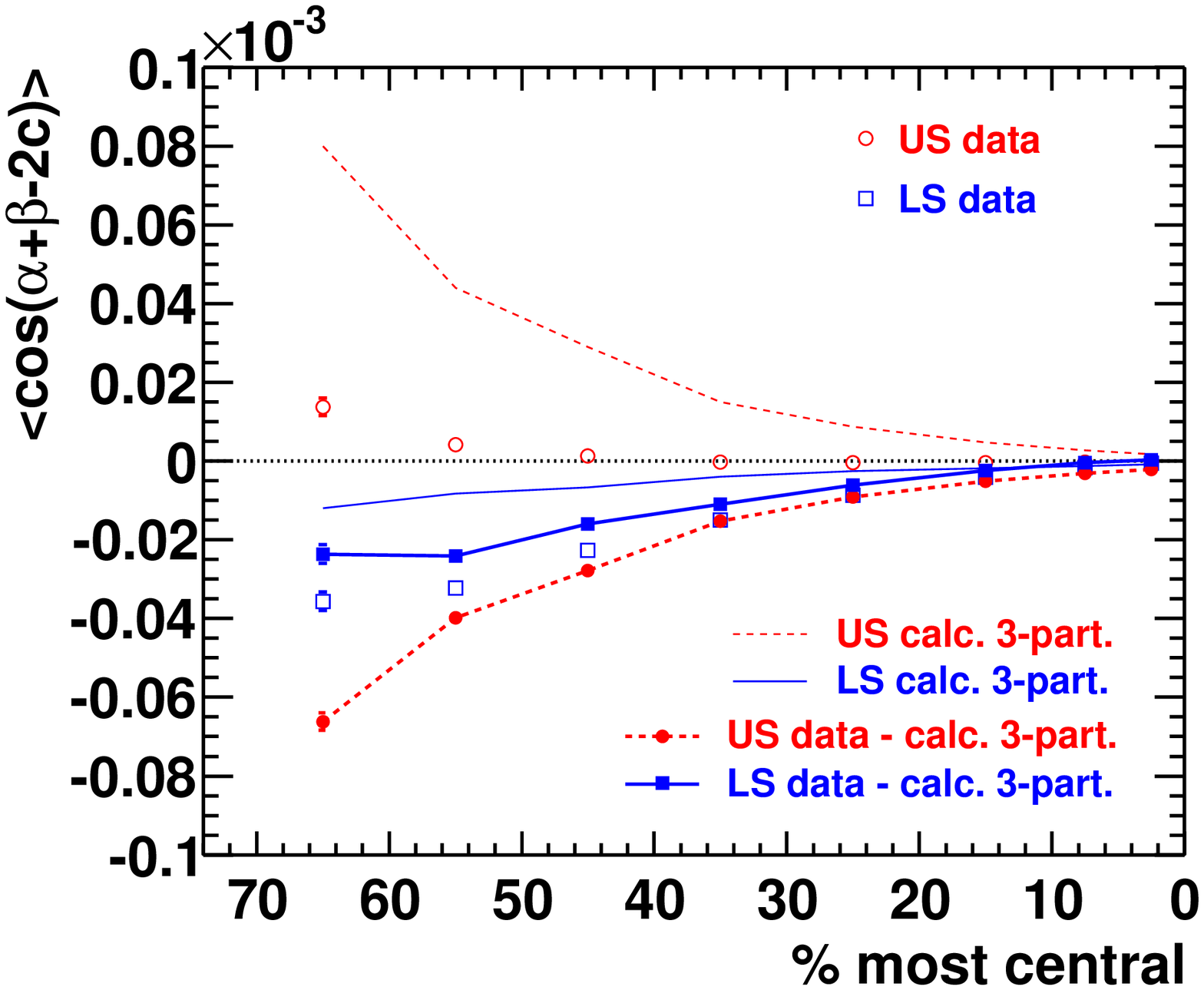}
\includegraphics[width=0.45\textwidth]{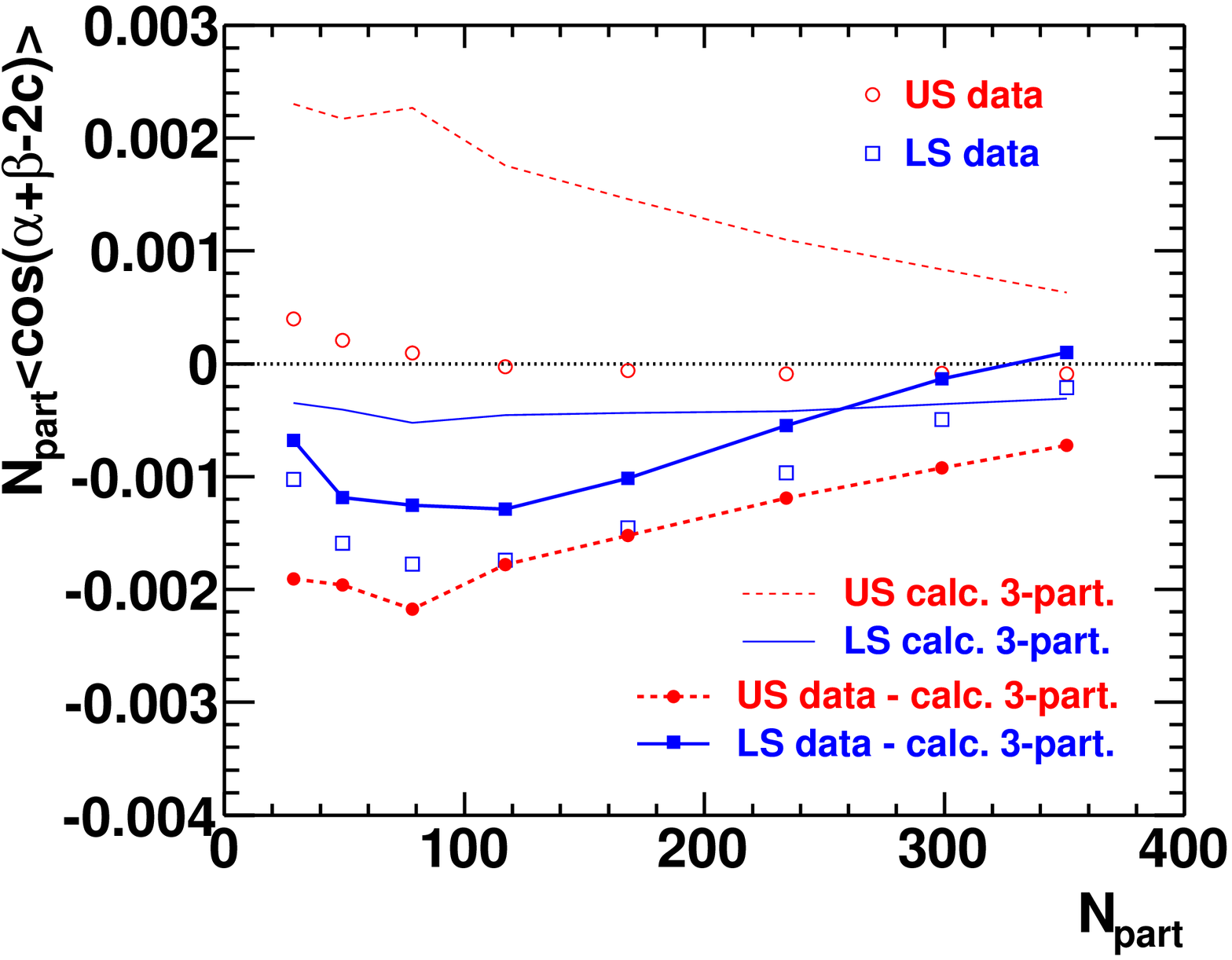}
\end{center}
\caption{(Color online) The measured three-particle correlators for US and LS particle pairs (open data points)~\cite{STAR}, the estimated three-particle correlation effects (thin lines) using inputs from measurements of two-particle angular correlations~\cite{Daugherity} and two-particle correlators~\cite{STAR}, and the remaining three-particle correlator magnitudes (solid data points), i.e. difference between open points and the thin lines. The left panel shows the three-particle correlators themselves versus centrality bin, and the right panel shows the number of participants ($\Npart$) scaled three-particle correlators versus $\Npart$.}
\label{fig2}
\end{figure}

Table~\ref{tab} shows the effect in three-particle azimuthal correlator estimated from the measured two-particle azimuthal correlator by Eqs.~(\ref{eq21a}) and (\ref{eq21b}), using cluster inputs from two-particle correlation measurements~\cite{Daugherity}. Figure~\ref{fig2} shows the estimated three-particle correlation effects in thin lines together with the measured three-particle correlator data in open points~\cite{STAR}. The estimated three-particle correlation effects from clusters are significantly larger than the US measurement $\mean{\cabc}_{_{US}}^{\meas}$~\cite{STAR}. This implies that there must be some cancellation of US correlation in the data from other effects (one candidate is two-particle correlation as we will discuss below). Those estimated for LS pairs are smaller than measurement $\mean{\cabc}_{_{LS}}^{\meas}$~\cite{STAR} in most of the centralities, by a factor of a few.

The three-particle correlation effects should be first removed from the three-particle correlator measurements. After removing the estimated three-particle effects, the three-particle correlators (from physics other than three-particle correlation) become the solid data points in Fig.~\ref{fig2}. As seen, both the LS and the US three-particle correlators are negative and seem to follow a regular trend. Those remaining effects can be due to two-particle correlation from clusters together with cluster anisotropies, as well as any other physics. 

\subsection{Effect of two-particle correlation from clusters}

We now discuss the effect of two-particle correlation. Our approach is to assume that the only remaining effects in the correlator measurements are the combined effects of two-particle correlations and cluster anisotropies, and see whether the cluster flow parameters extracted from such an approach are reasonable. If they are unreasonable, then there may be new physics, such as local strong parity violation. 

After removing three-particle correlation, the remaining three-particle correlator can now be factorized by Eq.~(\ref{eq1}) because particle $c$ is not correlated with particle pair $\alpha$ and $\beta$ through clusters. However, particle $c$ (hydro-like particle) can be still correlated with the cluster if clusters are anisotropic. Two-particle correlation from clusters together with cluster anisotropy can give non-zero contribution to the measured three-particle correlator.

We subtract the estimated three-particle correlation effects from the measured US and LS three-particle correlators, $\mean{\cabc}_{_{US}}^{\meas}$ and $\mean{\cabc}_{_{LS}}^{\meas}$, respectively, as shown in Fig.~\ref{fig2}. What is left in the three-particle correlator is two-particle correlation from clusters. Two-particle correlations can exist between $\alpha$ and $\beta$ (and $c$ is not from the cluster), or similarly between $\alpha$ and $c$, or between $\beta$ and $c$. The former is simply
\be
\mean{\cabc}_{\alpha\beta}=\mean{\cos(\dphi_\alpha+\dphi_\beta+2(\phicl-\psi)-2(c-\psi))}_{\alpha\beta}=\mean{\cos(\dphi_\alpha+\dphi_\beta)}_{\alpha\beta}\cdot\vcl\cdot\vc,
\label{eq21c}
\ee
while the latter, e.g. between $\alpha$ and $c$, is given by
\be
\mean{\cabc}_{\alpha c}=\mean{\cos(\dphi_\alpha-2\dphi_c-(\phicl-\psi)+(\beta-\psi))}_{\alpha c}=\mean{\cos(\dphi_\alpha-2\dphi_c)}_{\alpha c}\cdot v_{1,clust}\cdot v_{1,\beta}.
\label{eq21d}
\ee
Since direct flow $v_1$ is generally much smaller than elliptic flow $v_2$ at mid-rapidity, the latter correlations can be neglected and we shall only focus on two-particle correlation effect between $\alpha$ and $\beta$.
	
We divide the three-particle correlation corrected results by $\vc$ used in Ref.~\cite{STAR} to obtain $\mean{\cabp}_{_{US}}$ and $\mean{\cabp}_{_{LS}}$. The dilution factors are properly taken into account. Assuming the only remaining correlation is from clusters (no new physics), then
\bea
\mean{\cabp}_{_{US}}&=&\xUS\mean{\cabp}_{_{SA}}+(1-\xUS)\mean{\cabp}_{_{BB}},\label{eq22a}\\
\mean{\cabp}_{_{LS}}&=&\xLS\mean{\cabp}_{_{SA}}+(1-\xLS)\mean{\cabp}_{_{BB}},\label{eq22b}
\eea
where
\bea
\mean{\cabp}_{_{SA}}&=&\mean{\cos(\dphi_\alpha+\dphi_\beta+2\phicl-2\psi)}_{_{SA}}=\mean{\cab}_{_{SA}}\vSA=\wSA\vSA,\label{eq23a}\\
\mean{\cabp}_{_{BB}}&=&\mean{\cos(\dphi_\alpha+\dphi_\beta+2\phicl-2\psi)}_{_{BB}}=\mean{\cab}_{_{BB}}\vBB=-\wBB\vBB.\label{eq23b}
\eea
Here $\vSA$ and $\vBB$ are the elliptic flow parameters of SA- and BB-clusters weighted by the number of particle pairs per cluster (i.e. anisotropy of clusters with each cluster counted by the number of pairs in the cluster). In Eqs.~(\ref{eq23a}) and (\ref{eq23b}), we have assumed that the cluster two-particle azimuthal correlation shape is independent of the cluster orientation with respect to the reaction plane, so that we can factorize the cluster azimuthal spread and the cluster anisotropy. If particle azimuthal distribution in clusters depends on the cluster orientation, then the cluster anisotropy should be taken as an effective average anisotropy. From Eqs.~(\ref{eq22a}), (\ref{eq22b}), (\ref{eq23a}) and (\ref{eq23b}), we have
\bea
\mean{\cabp}_{_{US}}&=&\xUS\wSA\vSA-(1-\xUS)\wBB\vBB,\label{eq24a}\\
\mean{\cabp}_{_{LS}}&=&\xLS\wSA\vSA-(1-\xLS)\wBB\vBB.\label{eq24b}
\eea
From Eqs.~(\ref{eq24a}) and (\ref{eq24b}) we solve for $\vSA$ and $\vBB$:
\bea
\vSA&=&\frac{1}{\wSA}\frac{(1-\xLS)\mean{\cabp}_{_{US}}-(1-\xUS)\mean{\cabp}_{_{LS}}}{\xUS(1-\xLS)-\xLS(1-\xUS)},\label{eq25a}\\
\vBB&=&\frac{1}{\wBB}\frac{\xLS\mean{\cabp}_{_{US}}-\xUS\mean{\cabp}_{_{LS}}}{\xUS(1-\xLS)-\xLS(1-\xUS)}.\label{eq25b}
\eea

Our strategy now is to see what values of $\vSA$ and $\vBB$ are needed in order to reproduce the correlator measurements $\mean{\cabp}_{_{US}}$ and $\mean{\cabp}_{_{LS}}$  from STAR~\cite{STAR}, and judge whether the required cluster anisotropies are reasonable. Table I shows the obtained flow parameters $\vSA$ and $\vBB$ for SA- and BB-cluster particle pairs, respectively. The errors are propagated from statistical errors on the three-particle correlator measurements and 5\% errors on low-$\pT$ two-particle angular correlation measurements (same-side Gaussian amplitude, same-side Gaussian $\sigma$, and dipole amplitude). All errors are treated as uncorrelated. 

\begin{table}
\caption{Results from cluster model calculation. The $\xUS$ is the fraction of US pairs from SA-clusters; the $\xLS$ is that of LS pairs from SA-clusters; the $\xUS$ and $\xLS$ are obtained from $\mean{\cab}$ measurements by Eqs.~(\ref{eq14a}) and (\ref{eq14b}) using inputs from two-particle angular correlation measurements~\cite{Daugherity}, assuming no charge difference in BB-clusters (Eq.~(\ref{eq10})). Cluster $\mean{\cabc}_{\rm clust}$ is the calculated three-particle correlation effect assuming independent particle emission in clusters and binary collision scaling for the number of clusters. The $\vSA$ and $\vBB$ are the elliptic flow parameters by Eqs.~(\ref{eq25a}) and (\ref{eq25b}) to reproduce the three-particle correlator results measured by STAR~\cite{STAR}, after cluster three-particle effect removed; errors are propagated from statistical errors on the correlator measurements~\cite{STAR} and the assumed 5\% error on the two-particle angular correlation measurements~\cite{Daugherity}. The last two columns list the $\vSA$ and $\vBB$ parameters needed to reproduce the three-particle correlator results~\cite{STAR}, assuming vanishing three-particle correlation from clusters.}
\label{tab}
\begin{tabular}{c|cc|cccc|cc}
\hline
& & & \multicolumn{4}{c|}{Binary scaled clusters} & \multicolumn{2}{c}{Cluster three-particle}\\
& & & \multicolumn{2}{c}{$\mean{\cabc}_{\rm clust}$} & & & \multicolumn{2}{c}{correlation set to zero}\\
centrality & \hspace{2mm}$\xUS$\hspace{2mm} & \hspace{2mm}$\xLS$\hspace{2mm} & \hspace{8mm}US\hspace{8mm} & \hspace{8mm}LS\hspace{8mm} & \hspace{6mm}$\vSA$\hspace{6mm} & \hspace{6mm}$\vBB$\hspace{6mm} & \hspace{6mm}$\vSA$\hspace{6mm} & \hspace{6mm}$\vBB$\hspace{6mm} \\\hline
70-60\% & 0.85 & 0.36 & $8.0 \times 10^{-5}$ & $-1.2 \times 10^{-5}$ & $-0.20 \pm 0.09$ & $0.36 \pm 0.12$ & $0.23 \pm 0.02$ & $0.79 \pm 0.07$ \\
60-50\% & 0.78 & 0.42 & $4.4 \times 10^{-5}$ & $-8.2 \times 10^{-6}$ & $-0.10 \pm 0.05$ & $0.30 \pm 0.07$ & $0.22 \pm 0.01$ & $0.61 \pm 0.04$ \\
50-40\% & 0.70 & 0.44 & $2.9 \times 10^{-5}$ & $-6.7 \times 10^{-6}$ & $-0.10 \pm 0.04$ & $0.11 \pm 0.04$ & $0.19 \pm 0.01$ & $0.40 \pm 0.02$ \\
40-30\% & 0.66 & 0.43 & $1.5 \times 10^{-5}$ & $-3.9 \times 10^{-6}$ & $-0.05 \pm 0.02$ & $0.11 \pm 0.03$ & $0.16 \pm 0.01$ & $0.32 \pm 0.02$ \\
30-20\% & 0.62 & 0.43 & $8.7 \times 10^{-6}$ & $-2.6 \times 10^{-6}$ & $-0.05 \pm 0.02$ & $0.05 \pm 0.02$ & $0.14 \pm 0.01$ & $0.24 \pm 0.01$ \\
20-10\% & 0.61 & 0.42 & $4.7 \times 10^{-6}$ & $-1.8 \times 10^{-6}$ & $-0.07 \pm 0.01$ & $0.00 \pm 0.02$ & $0.10 \pm 0.01$ & $0.18 \pm 0.01$ \\
10-5\% & 0.61 & 0.41 & $2.8 \times 10^{-6}$ & $-1.2 \times 10^{-6}$ & $-0.13 \pm 0.02$ & $-0.08 \pm 0.02$ & $0.07 \pm 0.01$ & $0.12 \pm 0.01$ \\
5-0\% & 0.61 & 0.38 & $1.8 \times 10^{-6}$ & $-8.8 \times 10^{-7}$ & $-0.21 \pm 0.03$ & $-0.17 \pm 0.03$ & $0.03 \pm 0.01$ & $0.07 \pm 0.01$ \\
\hline
\end{tabular}
\end{table}

\begin{figure}
\begin{center}
\includegraphics[width=0.45\textwidth]{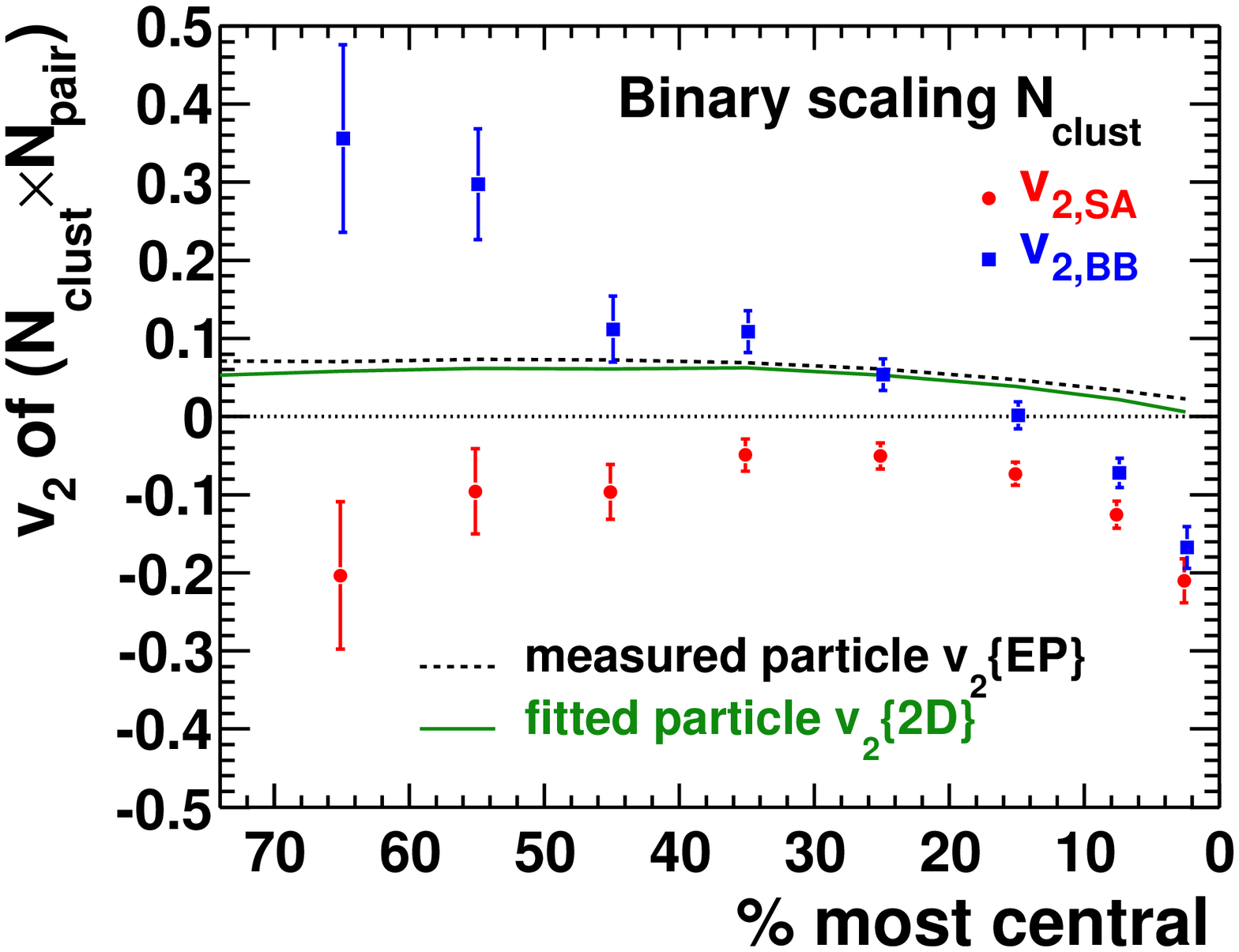}
\includegraphics[width=0.45\textwidth]{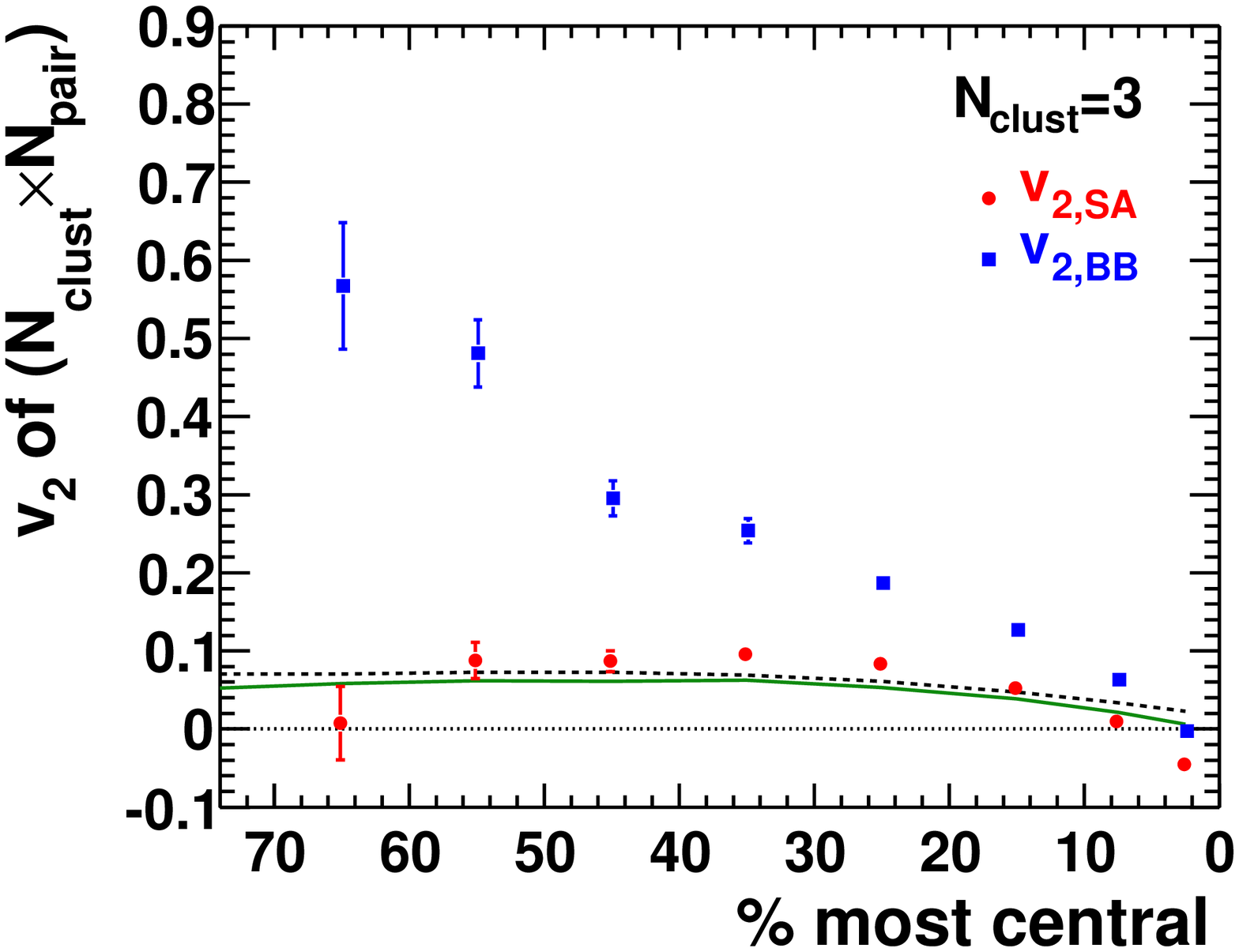}\\
\includegraphics[width=0.45\textwidth]{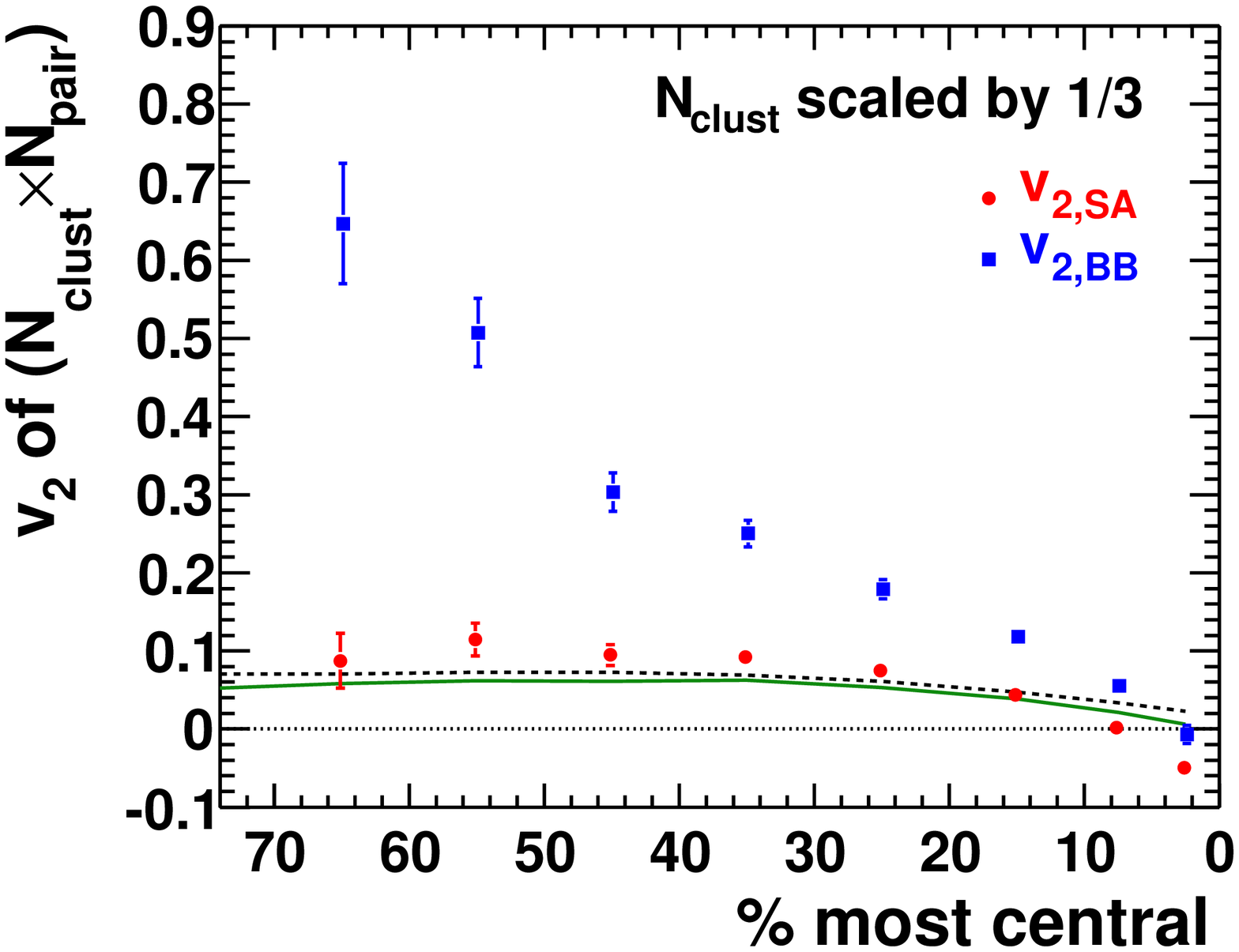}
\includegraphics[width=0.45\textwidth]{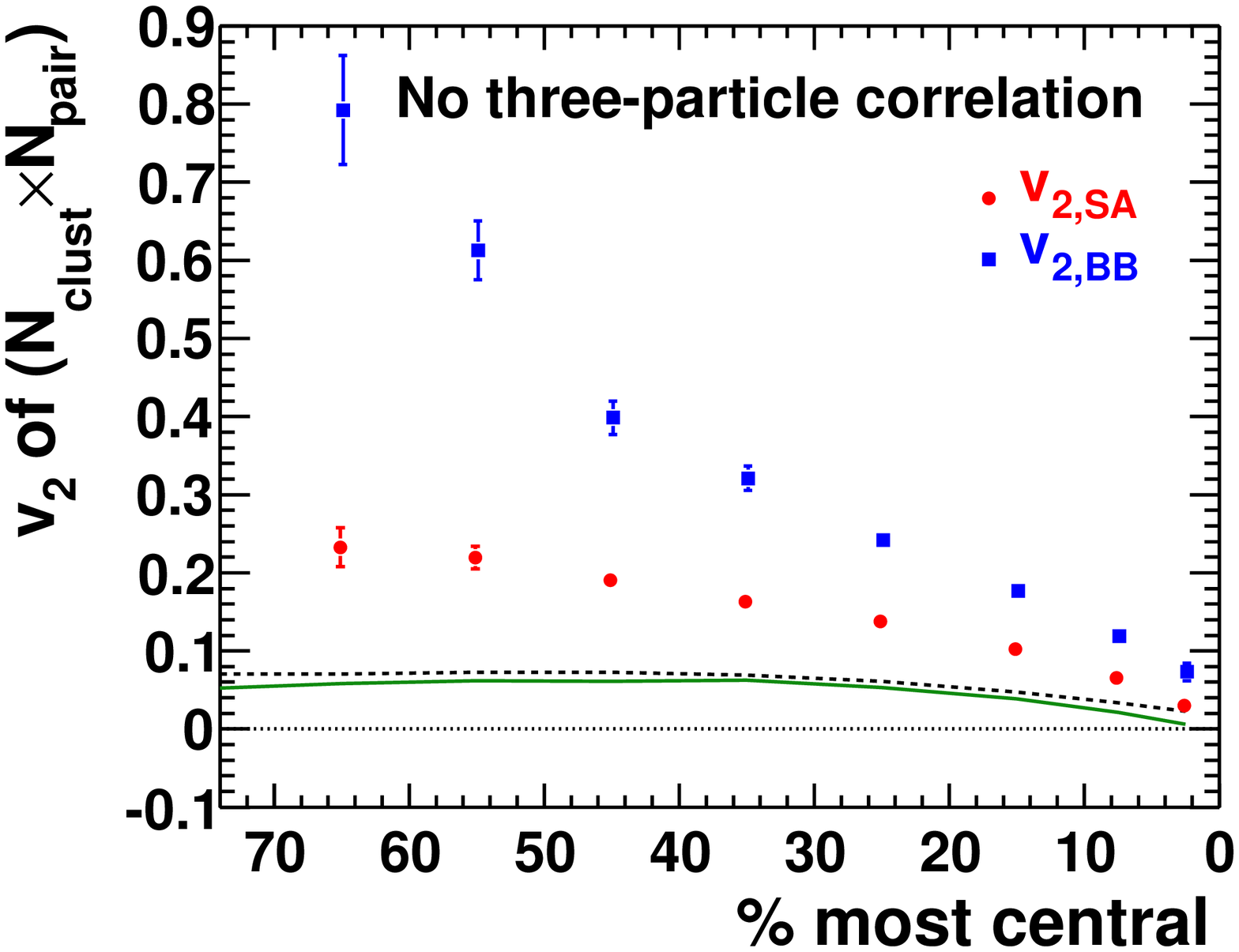}
\end{center}
\caption{(Color online) The $\vSA$ and $\vBB$ parameters (anisotropies of cluster population weighted by the number of particle pairs per cluster) required to reproduce the STAR three-particle correlator measurements~\cite{STAR} after removing cluster three-particle correlation effects calculated from two-particle angular correlation measurements~\cite{Daugherity} and two-particle correlator measurements~\cite{STAR}. The upper left panel assumes binary scaling for the number of clusters; the upper right panel assumes fixed cluster size of 3 particles; the lower left panel assumes that cluster size is a factor of 3 smaller than those in the upper left panel; and the lower right panel assumes a vanishing three-particle correlation from clusters. The particle elliptic flow measured by the event-plane method~\cite{flow} is shown in the dashed curve, and that from fit to two-particle correlation measurement~\cite{Daugherity,Kettler} is shown in the solid curve.}
\label{fig3}
\end{figure}

Figure~\ref{fig3} (upper left panel) depicts the obtained $\vSA$ and $\vBB$. The SA-cluster particle pairs $v_2$ is somewhat negative. Negative SA-cluster particle pair v2 is not impossible, perhaps even natural in the jet-quenching picture -- high-$\pT$ particles are suppressed more in the out-of-plane direction, generating more low-$\pT$ particles, thereby more SA-cluster particle pairs out-of-plane than in-plane. Positive anisotropy for BB-cluster particle pairs implies larger survival probability of BB-pairs in-plane than out-of-plane, again consistent with the jet-quenching picture. The magnitudes of the obtained $\vSA$ and $\vBB$ seem reasonable, however, the trends towards the most central collisions seem unreasonable. More discussions on the extracted flow parameters can be found in Section~\ref{sec3}.

\subsection{Dependence of results on model assumptions}

We have made two major assumptions in our study as we have italicized in Section~\ref{sec2A}:
\begin{itemize}
\item[(i)] {\em The number of clusters scales with binary collisions so we may estimate the cluster size;} and
\item[(ii)] {\em Particle emission azimuths within a cluster are independent of each other so we may factorize three-particle correlation as the product of two-particle correlations.}
\end{itemize}

Assumption (i) does not affect the two-particle dilution factor of Eqs.~(\ref{eq11a}) and (\ref{eq11b}) because the number of correlated particle pairs $\Ncl\Mcl^2$ is one of the measured quantities, as we noted already. However, the assumption does have an effect on the estimation of the number of correlated triplets $\Ncl\Mcl^3$, thereby on the three-particle dilution factor of Eqs.~(\ref{eq20a}) and (\ref{eq20b}). A larger $\Ncl$ than the binary scaling estimate would result in a smaller cluster size $\Mcl$, hence smaller three-particle correlation effect; a smaller $\Ncl$ would result in a larger three-particle effect. With binary scaling estimation of $\Ncl$, the cluster size varies in the range of $\Mcl\approx$~5-10 from peripheral to central Au+Au collisions, and the fraction of particles from clusters varies from 5-20\% of all particles measured in the final state~\cite{Daugherity,Wang}. Other studies, from multiplicity correlations, indicate that cluster size is only $\sim3$ whereas more particles originate from clusters~\cite{PHOBOSclust}. If cluster size is $\Mcl=3$ independent of centrality, then the fraction of particles from clusters would be 8-70\% of all particles from peripheral to central collisions and the number of clusters would scale more strongly than binary collisions. 

Assumption (ii) affects the factorization approximation in deriving the cluster three-particle correlation by Eqs.~(\ref{eq15a}) and (\ref{eq15b}). If independent emission of particles from clusters does not hold, then the three-particle cluster correlation result may not be valid. The factorization approximation also affects the connection between the two-particle correlations $\mean{\cos(\dphi_\alpha+\dphi_\beta)}=\mean{\cos(\dphi_\alpha-\dphi_\beta)}$ in Eqs.~(\ref{eq23a}) and (\ref{eq23b}).

To get a ``feeling'' about the effects of three-particle correlations from smaller clusters, we repeat our analysis by fixing the cluster size to be $\Mcl=3$, independent of centrality, while keeping the measured number of cluster particle pairs $\Ncl\Mcl^2$ the same. This changes the three-particle correlation effects from clusters. The cluster $\vSA$ and $\vBB$ that are needed to reproduce the measured three-particle correlators change accordingly. Figure~\ref{fig3} (upper right panel) shows the obtained $\vSA$ and $\vBB$. Both the $\vSA$ and $\vBB$ increase from the case in the upper left panel where binary collision scaling is assumed for the number of clusters resulting in large clusters. Furthermore, the values of $\vSA$ become mostly positive. The magnitude of $\vSA$ appears reasonable. The magnitude of $\vBB$ seems reasonable for most centralities except peripheral collisions where it is too large.

Besides the fixed cluster size, we have also tried reducing the cluster size by a constant factor of 3 for all centralities. The obtained $\vSA$ and $\vBB$ (which are needed to explain the three-particle correlator measurements) are shown in Fig.~\ref{fig3} (lower left panel). The results are similar to the previous results where cluster size is fixed to 3.

Reducing the cluster size reduces the effect of cluster three-particle correlation. The extreme would be to assume a vanishing three-particle correlation from clusters, and the only contribution to the measured three-particle correlator is the combined effect of two-particle correlation from clusters and cluster anisotropy. To test this extreme, we set the cluster three-particle correlation to zero, and repeat our analysis to extract the values of $\vSA$ and $\vBB$ that are needed to fully account for the measured three-particle correlators. The extracted $\vSA$ and $\vBB$ are tabulated in the last two columns of Table~\ref{tab}, and are shown in the lower right panel of Fig.~\ref{fig3}. Both the $\vSA$ and $\vBB$ are larger than the other cases, as expected from their trend with reducing effect from cluster three-particle correlation. Both $\vSA$ and $\vBB$ are positive, and $\vBB$ is larger than $\vSA$. The magnitude of $\vSA$ seems reasonable (see more discussion on this in Section~\ref{sec3}). The magnitude of $\vBB$ seems reasonable in central collisions, but is too large in peripheral collisions.

The signs and relative magnitudes of $\vSA$ and $\vBB$ in Fig.~\ref{fig3} (lower right panel) can be understood as follows. The US three-particle correlator is nearly zero. The three-particle correlators due to SA- and BB-cluster correlations should cancel each other. Because particle pairs from SA- and BB-clusters at the same location relative to the reaction plane give opposite sign three-particle correlators, the anisotropy of the SA- and BB-clusters have to be of the same sign (either positive or negative). Because more US pairs come from SA-clusters than BB-clusters, the anisotropy of BB-cluster particle pairs has to be larger than those of SA-clusters in order to have the averaged three-particle correlator to be nearly zero. The sign of the LS three-particle correlator is decided by BB-clusters, because more LS pairs come from BB-clusters and $\vBB$ is larger than $\vSA$ in absolute magnitude. BB-cluster particle pairs in-plane gives negative $\mean{\cabp}$ and those out-of-plane gives positive $\mean{\cabp}$. In order for the final averaged LS pair correlator $\mean{\cabp}_{_{LS}}$  to be negative, the $\vBB$ has to be positive.

\section{Discussion on Cluster Particle Pair Anisotropy\label{sec3}}

The cluster anisotropy $\vSA$ and $\vBB$ obtained above are the magnitudes of modulation of the number of clusters weighted by the average number of particle pairs per cluster. One way to gauge whether the obtained $\vSA$ and $\vBB$ are reasonable or not is to see how large the hydro-like particle $\vbg$ has to be in order, together with cluster $v_2$, to reproduce the observed final state particle $v_2^{\meas}$. Since our clusters are large, having on average 5-10 particles, the particles from back-to-back pairs are part of those from SA-clusters. We can therefore consider only SA-cluster pair $\vSA$. Note this may not be accurate in peripheral collisions where a BB-cluster of a single back-to-back particle pair is not part of a SA-cluster pair, thus the BB $\vBB$ has to be also considered. But for our purpose, it is sufficient to use only SA $\vSA$ to get a ``feeling''. 

The hydro-like particle $\vbg$ can be obtained from the following two scenarios.
\begin{itemize}
\item Assuming cluster size does not vary with respect to the reaction plane, then cluster $\vSA$ translates directly into cluster particle $v_2$ by 
\be
\vSA^{\rm particle}=\wwSA\vSA,\label{eq26}
\ee
where $\wwSA$ is the angular spread of cluster particles relative to the cluster axis by Eq.~(\ref{eq16}). The hydro-like particle $\vbg$ is then given by
\be
(1-f)\vbg+f\wwSA\vSA=v_2^{\meas},\label{eq27}
\ee
where $f$ is the fraction of particles from clusters.
\item Assuming cluster orientation is isotropic (e.g. initial hard scattering products), but cluster size varies from in-plane to out-of-plane, then approximately half of the cluster $\vSA$ translates into cluster particle $v_2$ by
\be
\vSA^{\rm particle}=\wwSA\vSA/2.\label{eq28}
\ee
The hydro-like particle $\vbg$ is then given by
\be
(1-f)\vbg+f\wwSA\vSA/2=v_2^{\meas}.\label{eq29}
\ee
\end{itemize}

We use the $v_2$\{{\sc 2d}\} fit to the two-particle angular correlation data~\cite{Daugherity} as the measured $v_2^{\meas}$, because the non-flow same-side correlation peak (part of the cluster pair correlation) should be excluded. In fact, if the fit model~\cite{Daugherity} used to separate elliptic flow and non-flow correlations is accurate, then the fit $v_2$\{{\sc 2d}\} should be the true elliptic flow (correlation related to the reaction plane)~\cite{Wang}. Note that the true elliptic flow may not be necessarily equal to the hydro-like elliptic flow, but is the net sum of the hydro-like elliptic flow and the product of cluster correlation and cluster elliptic flow~\cite{Wang}, as shown in Eqs.~(\ref{eq27}) and (\ref{eq29}).

\begin{figure}
\begin{center}
\includegraphics[width=0.45\textwidth]{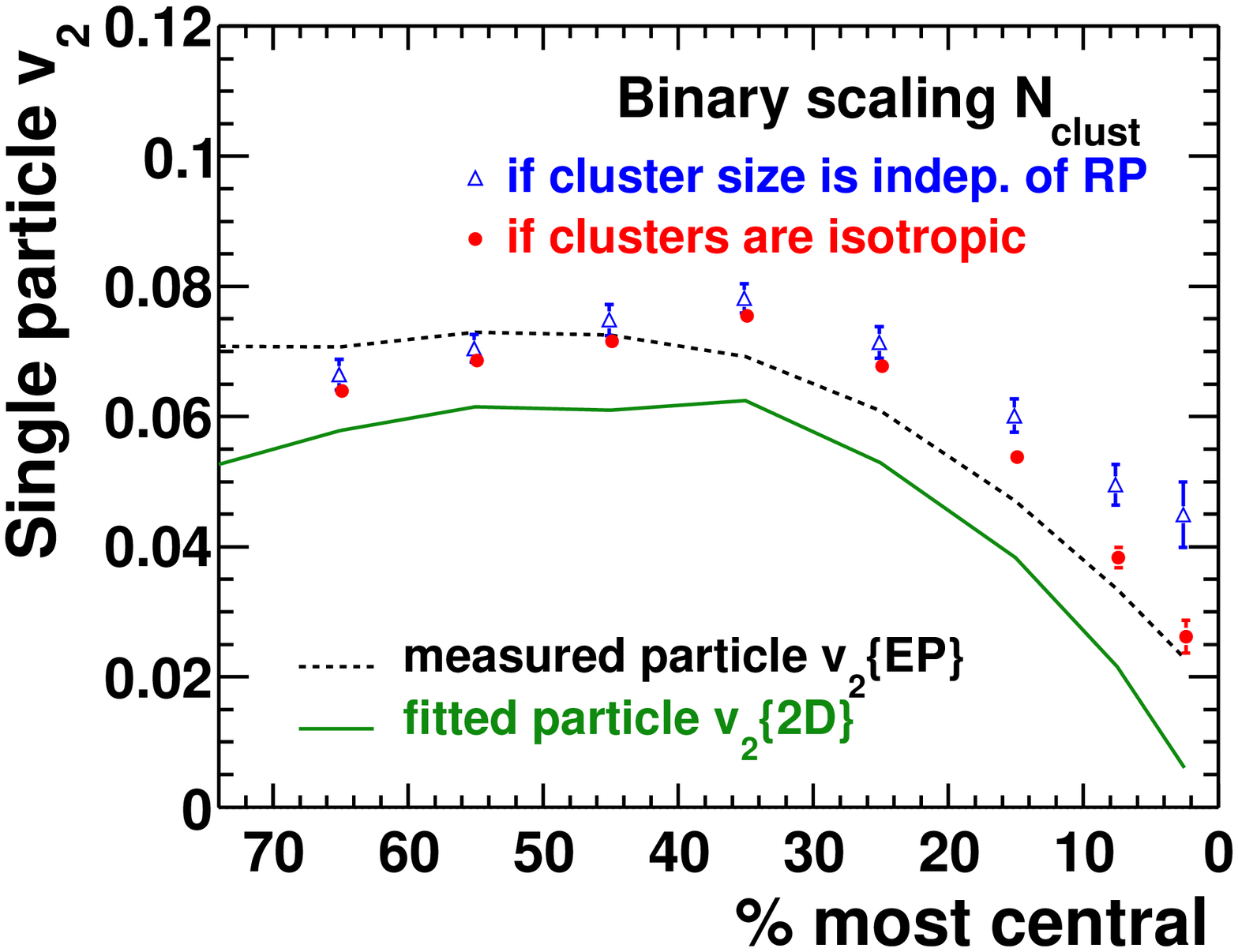}
\includegraphics[width=0.45\textwidth]{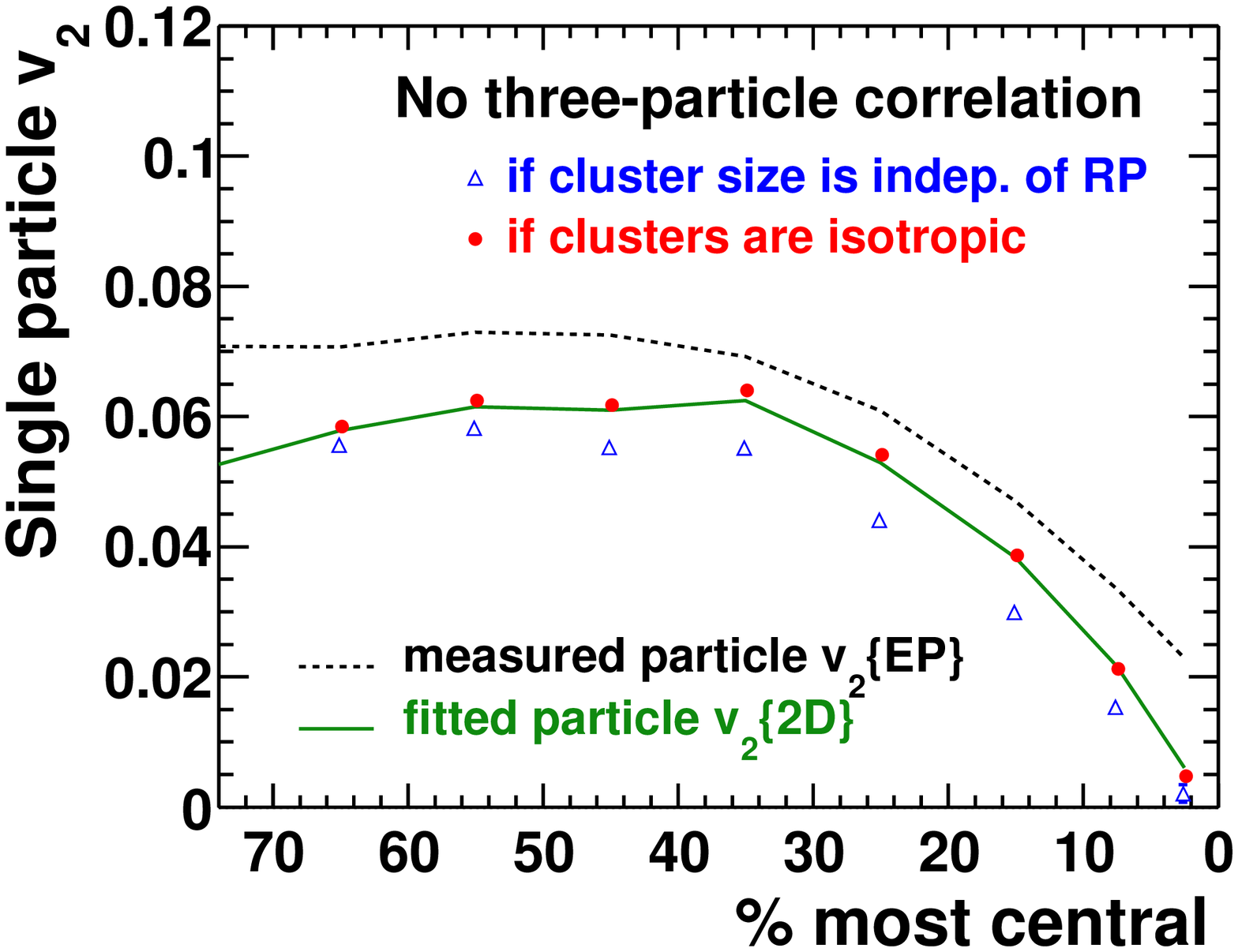}
\end{center}
\caption{(Color online) hydro-like particle particle $\vbg$ that is needed in order, together with the estimated $\vSA$, to reproduce the fit $v_2$\{{\sc 2d}\}. The left panel corresponds to $\vSA$ in Fig.~\ref{fig3} upper left panel where the number of clusters is assumed to scale with binary collisions. The right panel corresponds to $\vSA$ in Fig.~\ref{fig3} lower right panel where the three-particle correlation from clusters is set to zero. The fraction of cluster particles is kept as same as that for the left panel.}
\label{fig4}
\end{figure}

Figure~\ref{fig4} shows the obtained hydro-like particle $\vbg$ from Eqs.~(\ref{eq27}) and (\ref{eq29}) together with the $v_2$\{{\sc 2d}\}. The $v_2$\{{\sc ep}\} measured by the event-plane method, which contains a large contribution from non-flow, is also shown. Its difference from $v_2$\{{\sc 2d}\} gives a good estimate of uncertainty of all available elliptic flow measurements. The left panel of Fig.~\ref{fig4} shows the case where the number of clusters scales with binary collisions corresponding to the upper left panel of Fig.~\ref{fig3}. The obtained hydro-like particle $v_2$ is larger than the measured particle $v_2$ to account for the negative cluster $\vSA$, but not much larger. The $\vbg$ values seem reasonable, suggesting that the $\vSA$ and $\vBB$ may be reasonable too. 

The right panel of Fig.~\ref{fig4} shows the case where the three-particle correlation from clusters is set to zero, corresponding to the lower right panel of Fig.~\ref{fig3}. Although the fraction of particles from clusters does not matter for this case because the three-particle correlation from clusters is set to zero, we need the fraction to obtain the hydro-like particle $\vbg$ by Eqs.~(\ref{eq27}) and (\ref{eq29}). We use the same fraction of particles from clusters as in the above case (assuming binary collision scaling for the number of clusters). The calculated hydro-like particle $\vbg$ is smaller than the measured particle $v_2$ for cluster size independent of the reaction plane to offset the larger cluster particle $\vSA^{\rm particle}$ (by Eq.~(\ref{eq26})). For isotropic clusters, the hydro-like particle $\vbg$ is not much different from the measured particle $v_2$ because the cluster particle $\vSA^{\rm particle}$ due to cluster anisotropy (by Eq.~(\ref{eq28})) approximately equal to the measured particle $v_2$. The hydro-like particle $\vbg$ is reasonable for this case too, again suggesting that the $\vSA$ and $\vBB$ required to explain the three-particle correlator measurements are reasonable.

\section{Conclusions}

Cluster model parameters are calculated by Eqs.~(\ref{eq5}), (\ref{eq6}), and (\ref{eq16}) using two-particle angular correlation data~\cite{Daugherity} and two-particle azimuth correlator $\mean{\cab}$ measurements from STAR~\cite{STAR}. Fractions of unlike-sign (US) and like-sign (LS) particle pairs from small-angle (SA) and back-to-back (BB) clusters are obtained by Eqs.~(\ref{eq14a}) and (\ref{eq14b}) assuming no charge difference in BB two-particle correlation. Three-particle correlation effects are estimated by Eqs.~(\ref{eq21a}) and (\ref{eq21b}) assuming independent emission of particles in clusters, and are positive for SA-clusters and negative for BB-clusters. The estimated three-particle effects are removed from the three-particle azimuth correlator $\mean{\cabc}$ measurements~\cite{STAR}. The remaining correlator magnitudes of $\mean{\cabp}=\mean{\cabc}/\vc$ are assumed to come entirely from cluster two-particle correlations (i.e. no new physics), and are used to determine elliptic flow parameters of SA- and BB-clusters by Eqs.~(\ref{eq25a}) and (\ref{eq25b}). 

Cluster size is not measured. A wide range of assumptions are made, ranging from binary collision scaling of cluster abundance, resulting in large cluster three-particle correlation, to zero cluster size yielding vanishing three-particle correlation effect. These assumptions do not affect the cluster two-particle correlation which is constrained by the two-particle angular correlation measurements~\cite{Daugherity}. The magnitudes of the obtained cluster anisotropy (azimuthal modulation in the number of clusters weighted by the number of particle pairs per cluster) to fully account for the three-particle correlator measurements~\cite{STAR} seem reasonable, except for peripheral collisions. The hydro-like particle flow magnitude, in order to make up to the measured inclusive particle flow together with the cluster particle flow, appears reasonable too. 

Cluster particle correlations may originate from (semi-)hard scatterings. It is therefore natural to expect that cluster effect would increase with $\pT$, although the $\pT$-dependence is not studied in this paper due to the lack of $\pT$-dependent measurements of two-particle angular correlations. It is worth to note that the measured azimuth correlators indeed increase with $\pT$~\cite{STAR}, an observation that cannot be explained by local strong parity violation but may be expected from cluster particle correlations.

In conclusion, our results from {\em conventional} physics of cluster particle correlations suggest that no new physics is {\em required} to explain the two- and three-particle azimuth correlator measurements by STAR. Our conclusion is complementary to the experimental findings in Ref.~\cite{STAR} that the azimuth correlator data in itself does not allow to conclude on the local strong parity violation.

\section*{Acknowledgments}

I thank my STAR collaborators for useful discussions. 
This work is supported by U.S. Department of Energy under Grant DE-FG02-88ER40412.


\end{document}